\documentclass[12pt]{iopart}

\usepackage{graphicx}
\begin{document}

\title[Surface tension in an intrinsic curvature model]{Surface tension in an intrinsic curvature model with fixed one-dimensional boundaries}

\author{Hiroshi Koibuchi}

\address{Department of Mechanical and Systems Engineering, Ibaraki National College of Technology, Nakane 866 Hitachinaka, Ibaraki 312-8508, Japan}
\ead{koibuchi@mech.ibaraki-ct.ac.jp}

\begin{abstract}
A triangulated fixed connectivity surface model is investigated by using the Monte Carlo simulation technique. In order to have the macroscopic surface tension $\tau$, the vertices on the one-dimensional boundaries are fixed as the edges (=circles) of the tubular surface in the simulations. The size of the tubular surface is chosen such that the projected area becomes the regular square of area $A$. An intrinsic curvature energy with a microscopic bending rigidity $b$ is included in the Hamiltonian. We found that the model undergoes a first-order transition of surface fluctuations at finite $b$, where the surface tension $\tau$ discontinuously changes. The gap of $\tau$ remains constant at the transition point in a certain range of values $A/N^\prime$ at sufficiently large $N^\prime$, which is the total number of vertices excluding the fixed vertices on the boundaries. The value of $\tau$ remains almost zero in the wrinkled phase at the transition point while $\tau$ remains negative finite in the smooth phase in that range of $A/N^\prime$.

\end{abstract}

\maketitle

\section{Introduction}\label{intro}
Membranes are known to have a large variety of shapes, which change from one to the other depending on some external conditions such as mechanical conditions or some chemical environments \cite{SEIFERT-LECTURE2004,Yoshikawa-FEBS2003,Hotani-JMB1984}. One interesting topic in membrane physics is to clarify how the membrane shape is maintained and to understand the shape of membranes in terms of the notion of phase transitions \cite{NELSON-SMMS2004-1,GK-SMMS2004,Bowick-PREP2001,Gompper-Schick-PTC-1994}. For this reason, the conventional surface model of Helfrich and Polyakov \cite{HELFRICH-1973,POLYAKOV-NPB1986,KLEINERT-PLB1986} has been investigated so far by many groups \cite{Peliti-Leibler-PRL1985,DavidGuitter-EPL1988,PKN-PRL1988,KANTOR-NELSON-PRA1987,WHEATER-NPB1996}. Including the conventional model we have currently a variety of surface models; extrinsic or intrinsic curvature models \cite{KD-PRE2002,KOIB-PRE-20045-NPB-2006,KOIB-EPJB2004,KOIB-PLA2005,KOIB-PLA2006,KOIB-PRE2004,KOIB-EPJB2007-2}, two-dimensional curvature or one-dimensional curvature models \cite{KOIB-PRE2007,KOIB-EPJB2007-3,KOIB-PLA2007}, fixed connectivity models or fluid surface models. 

Phase transitions in the surface models are driven by thermal fluctuations and are dependent mainly on the temperature $T$, because the bending rigidity $b[kT]$, which is a parameter included in the Hamiltonian, has the unit of $kT$. Since $b$ is a microscopic parameter, the detailed information of the dependence of $b$ on $T$ is unknown, however, $b$ is expected to vary with $T$. 

The phase transition can also be influenced by the boundary conditions, which impose a constraint on the dynamical variable $X$ the position of the surface. The shape of membranes depends just on the value of $X$, and hence the phase structure crucially depends on the external conditions that place a constraint on $X$. A flow field is expected to influence the phase structure of the surface model \cite{NG-PRL2004}.

 An example of the boundary conditions is to fix a part of the surface. Fixing two vertices separated by a distance $L$ on a spherical surface, we have a surface model with such nontrivial constraints on $X$.  This model leads us to calculate the macroscopic string tension $\sigma$ of the surface by equating $\exp (-\sigma L) \sim Z$ at sufficiently large $L$, where $Z$ is the partition function of the surface model \cite{AMBJORN-NPB1993,WHEATER-JP1994}. The expected scaling relation of $\sigma$ with respect to $N$ is observed, where $N$ is the total number of vertices of the triangulated surface \cite{KOIB-PLA2004-2,KOIB-EPJB2005,KOIB-JSAT2006-1,KOIB-JSAT2006-2}.

 It is also possible to fix the tubular surface with one-dimensional boundaries of area $A$. Then, the macroscopic surface tension $\tau$ is expected to be obtained by $\exp (-\tau A) \sim Z$ at sufficiently large $A$. Therefore, it is interesting to study a surface model that has a phase transition and see how the surface tension reflects the phase transition. However, $\tau$ was reported only in \cite{AMBJORN-NPB1993}. 
 
In this paper, we study an intrinsic curvature model with two one-dimensional boundaries, which are a couple of circles separated by $L$. Therefore, the surface spans a cylindrical tube from one circle to the other. The reason why we use a surface with two boundaries is because it seems more convenient to use a surface with two boundaries than a surface with one boundary for experimental measurements of the surface tension. Let $R$ be the diameter of the boundary circles, then the projected area $A$ is given by $RL$, which is proportional to the area $\pi RL$ of the tube. In this paper, $R$ is chosen to be $R\!=\!L$, then we have $A\!=\!L^2$. The area $\pi RL$ of the tube can also be used for the calculation of $\tau$, however, we use the projected area $L^2$ for simplicity. We should emphasize that it is nontrivial whether the model undergoes a phase transition because the surface is fixed by one-dimensional boundaries, which include many vertices, and the size of the boundaries increases with the system size, and therefore, the phase structure of the model without boundaries \cite{KOIB-EPJB2004} or with point boundaries \cite{KOIB-JSAT2006-1} is expected be influenced by the one-dimensional boundaries. The purpose of the study is to see whether or not the model undergoes phase transitions, and moreover, to see how the transition is reflected in the surface tension if the transition occurs in the model. 

\section{Model and Monte Carlo technique}
\begin{figure}[hbt]
\centering
\includegraphics[width=8.0cm]{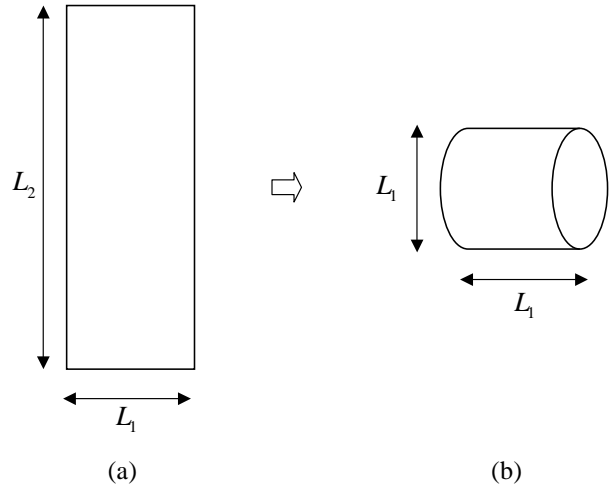}
\caption{(a) A rectangular surface of size $L_1 \times L_2$, and (b) a tubular surface of length $L_1$ and diameter $L_1$, where $L_2\!=\!\pi L_1$. The projected area of the tube in (b) is given by $A\!=\!L_1^2$.}
\label{fig-1}
\end{figure}
Firstly, we show how to construct the triangulated surface on which the model is defined. Figure \ref{fig-1}(a) shows a rectangle of size $(L_1, L_2)$. Then we obtain a tubular surface in Fig. \ref{fig-1}(b) from the rectangle in Fig. \ref{fig-1}(a) by bending and sewing up a couple of sides of the rectangle. The length $L_2$ of the rectangle in Fig. \ref{fig-1}(a) is chosen to be $L_2=\pi L_1$ so that the diameter $R$ of the cylinder in Fig. \ref{fig-1}(b) satisfies $R\!=\!L_1$. The rectangle in Fig. \ref{fig-1}(a) and hence the cylindrical surface in Fig. \ref{fig-1}(b) can be regularly triangulated, and then we have a triangulated open tubular surface with a couple of one-dimensional boundaries. We should note that the surface tension $\tau$ is obtained by a surface with a single one-dimensional boundary such as the rectangle in Fig. \ref{fig-1}(a) at sufficiently large area $L_1L_2$. However, we should remind ourselves of that it is also possible to define $\tau$ on a tubular surface such as the one in Fig. \ref{fig-1}(b).

\begin{figure}[hbt]
\centering
\includegraphics[width=9.5cm]{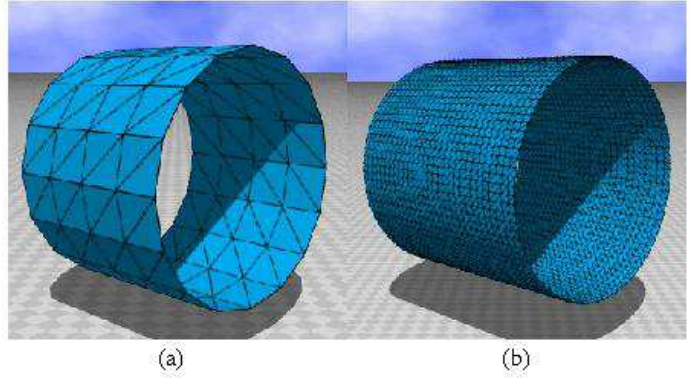}
\caption{Snapshots of lattices of size (a) $N\!=\!80$, where $(L_1,L_2)\!=\!(5,16)$ and (b) $N\!=\!2820$, where $(L_1,L_2)\!=\!(30,94)$. }
\label{fig-2}
\end{figure}
 Figures \ref{fig-2}(a) and \ref{fig-2}(a) show a triangulated surface of size $(L_1,L_2)\!=\!(5,16)$ and that of  $(L_1,L_2)\!=\!(30,94)$. Because $L_1$ and $L_2$ are integers, $L_2$ is given by $L_2\!=\![\pi L_1]$, where $[x]$ denotes the integer obtained by rounding $x$. The coordination number $q$ is $q\!=\!6$ except at the boundary vertices; this can be seen in the lattice in  Fig. \ref{fig-2}(a). We should note that $L_1\!=\!5$ in Fig. \ref{fig-2}(a) denotes the total number of vertices along the axis of the tube and hence deviates from the length $L\!=\!4$ by one. The final numerical results depend only on the total number of vertices and on the reduced projected area, which will be introduced below. 

We multiply $L_1$ by a scale factor $r$ in order to vary the projected area $A$ such that $A\!=\!(r L_1)^2$.  Table \ref{table-1} shows the numbers $N$, $N^\prime$, $L_1$, $L_2$, and the projected areas $A_1$, $A_2$, and $A_3$, and the factors $r$, which characterize the triangulated tubes used in the simulations. $N^\prime(\!=\!N\!-\!2L_2)$ in Table \ref{table-1} is the total number of internal vertices. The total number of vertices $N(=\!L_1L_2)$ includes  $2L_2$ the total number of the boundary vertices. Three types of the projected areas $A_i(i=1,2,3)$ are assumed for the surfaces of size $N\!=\!2820$ and  $N\!=\!5040$, and two types of $A_i(i=1,2)$ are assumed for the $N\!=\!9460$ and  $N\!=\!17550$ surfaces. The scale factor $r_i$, which is given by $r_i\!=\!\sqrt{A_i}/L_1$, is as follows: $r_1\!=\!17/75\!\simeq\!0.227$ and  $r_2\!=\!19/75\!\simeq\!0.253$ for $A\!=\!289$ and  $A\!=\!361$ on the $N\!=\!17550$ surface, respectively. 
\begin{table}[hbt]
\caption{ The numbers that characterize the triangulated tubes for the simulations; the total number of vertices $N$ including the boundary vertices, the total number of internal vertices $N^\prime$, the length $L_1$ of the cylinder, the circumference $L_2$ of the cylinder, and the projected areas $A_1$, $A_2$, and $A_3$, and the scale factors $r_i\!=\!\sqrt{A_i}/L_1$. \label{table-1}}
\begin{center}
 \begin{tabular}{ccccccccccc}
$N$  & $N^\prime$  & $L_1$ & $L_2$  && $A_1$ & $A_2$ & $A_3$ & $r_1$ & $r_2$ & $r_3$\\
 \hline
  2820   &  2632   & 30  &  94  && 100 & 144 & 196 & 0.333 & 0.4   & 0.467  \\
  5040   &  4788   & 40  &  126 && 144 & 196 & 256 & 0.3   & 0.35  & 0.4    \\
  9460   &  9116   & 55  &  172 && 196 & 256 &     & 0.255 & 0.291 &    \\
 17550   & 17082   & 75  &  234 && 289 & 361 &     & 0.227 & 0.253 &    \\
 \hline
 \end{tabular}
\end{center}
\end{table}

The model is defined by the Hamiltonian, which is a linear combination of the Gaussian bond potential $S_1$ and the intrinsic curvature energy $S_2$ such that
\begin{equation}
\label{S1S2}
S=S_1+b S_2, \quad S_1=\sum_{(ij)} (X_i-X_j)^2,\quad S_2=\sum_{i} (\delta_i - \Delta_i )^2,
\end{equation}
where $\delta_i$ is defined by the summation over the vertex angle of the triangles meeting at the vertex $i$. The symbol $\Delta_i$ is defined by
\begin{equation}
\label{delta}
 \Delta_i = \cases{
                     2\pi & (i = {\rm internal\; vertex}), \\
                     \pi  & (i = {\rm boundary\; vertex}). \\
                   }
\end{equation}
On the boundaries $\Delta_i$ is fixed to $\Delta_i\!=\!\pi$, while it is fixed to  $\Delta_i\!=\!2\pi$ at the internal vertices. 

The partition function $Z$ of the model is defined by 
\begin{equation} 
\label{Part-Func}
 Z =\int \prod _{i=1}^{N^\prime} d X_i \exp\left[-S(X)\right],
\end{equation} 
where $X_i (\in {\bf R}^3)$ is the position of the internal vertex $i$ on the triangulated surface. The multiple integrations in $Z$ are performed by leaving the boundary vertices fixed.

The surface tension $\tau$ of the model is given by
\begin{equation} 
\label{surfacetension}
\tau = {2\langle S_1\rangle - 3N^\prime \over 2A},
\end{equation} 
where $A$ is the projected area of the surface and is given by $A\!=\!(rL_1)^2$ denoted as above. The expression of $\tau$ in Eq.(\ref{surfacetension}) is obtained by the scale invariant property of the partition function \cite{AMBJORN-NPB1993,WHEATER-JP1994}.

The canonical Monte Carlo (MC) simulation technique is adopted to simulate the multiple integrations $\prod _i d X_i$ with the Boltzmann weight $\exp\left[-S(X)\right]$ in the partition function. The vertex position $X$ is shifted to a new position $X^\prime\!=\!X\!+\!{\mit \Delta}X$ with a three-dimensional small random vector ${\mit \Delta}X$. The new position $X^\prime$ is accepted with the probability ${\rm Min}[1,\exp({\mit \Delta}S)]$, where ${\mit \Delta}S\!=\!S({\rm new})\!-\!S({\rm old})$. 

\section{Results}
\begin{figure}[hbt]
\centering
\includegraphics[width=8.5cm]{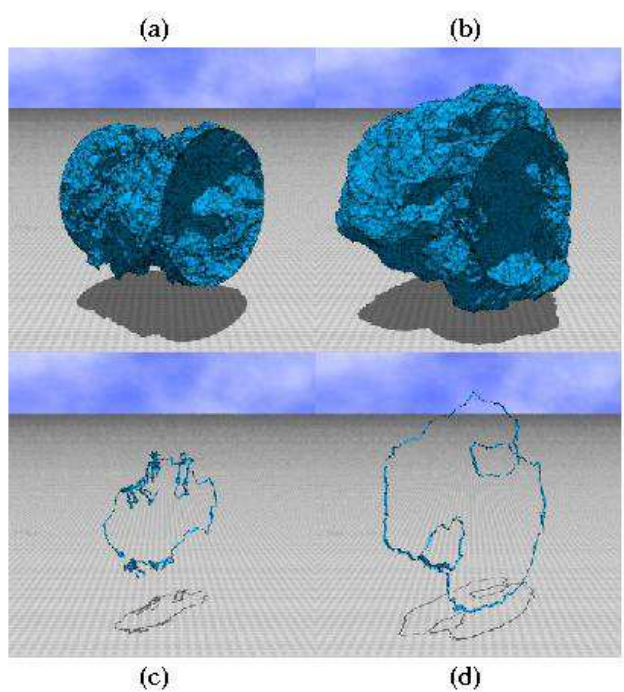}
\caption{Snapshots of surfaces of size $N\!=\!17550$ obtained  at (a) $b\!=\!11$ (wrinkled phase) and at (b) $b\!=\!12$ (smooth phase), and (c), (d) the corresponding surface sections. The figures are drawn in the same scale. }
\label{fig-3}
\end{figure}
We show snapshots of surfaces in Figs. \ref{fig-3}(a) and \ref{fig-3}(b) and the surface sections in Figs. \ref{fig-3}(c) and \ref{fig-3}(d), where the surface size is $N\!=\!17550$. The snapshots in Figs. \ref{fig-3}(a) and \ref{fig-3}(c) are obtained in the wrinkled phase at $b\!=\!11$, and those in Figs. \ref{fig-3}(b) and \ref{fig-3}(d) are in the smooth phase at $b\!=\!12$. Both surfaces are corresponding to the projected area $A\!=\!361$. We find that the surfaces are spanning from one boundary to the other boundary and do not shrink to one-dimensional object, and therefore the surface tension $\tau$ is expected to be evaluated in both phases.

\begin{figure}[hbt]
\centering
\includegraphics[width=14.5cm]{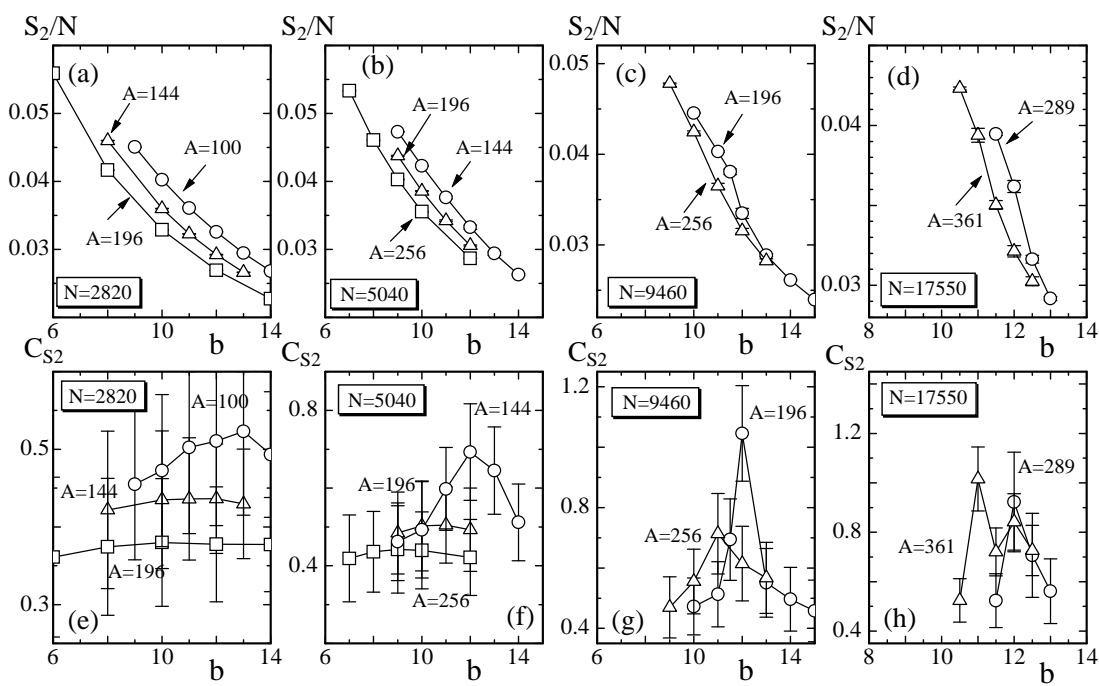}
\caption{The intrinsic curvature energy $S_2/N$ vs. $b$ obtained on the surfaces of size (a) $N\!=\!2820$, (b) $N\!=\!5040$, (c) $N\!=\!9460$, and (d) $N\!=\!17550$, and (e)--(h) the specific heat $C_{S_2}$ vs. $b$ on the same sized surfaces. The error bars denote the standard errors.}
\label{fig-4}
\end{figure}
Figures \ref{fig-4}(a) -- \ref{fig-4}(d) show the intrinsic curvature energy $S_2/N$ vs. $b$, which are obtained on the surfaces of size $N\!=\!2820$, $N\!=\!5040$, $N\!=\!9460$, and $N\!=\!17550$, respectively. The corresponding specific heat $C_{S_2}$ is defined by
\begin{equation}
\label{fluctuation-S2}
C_{S_2} = {b^2\over N} \langle \; \left( S_2 \!-\! \langle S_2 \rangle\right)^2\rangle,
\end{equation}
and it is shown in Figs. \ref{fig-4}(e) -- \ref{fig-4}(h). We see in  $C_{S_2}$ an anomalous peak, which reflects a phase transition, where "anomalous peak" denotes that the peak value increases with increasing $N$. We see that the peak value remains low in the case of the largest surface of $N\!=\!17550$ with $A\!=\!289$. This is because the values of $b$, assumed in the simulations, are out of the transition region, which is narrow in such a large surface. The transition appears to be of first-order, because $S_2/N$ discontinuously changes on the surfaces of $N\!=\!9460$ with $A\!=\!196$ and $N\!=\!17550$ with $A\!=\!289$ in Figs. \ref{fig-4}(c) and \ref{fig-4}(d). 

We must comment on the value $S_2/N$, where $N$ includes the boundary vertices, on which the definition of $S_2$ is slightly different from that on the internal vertices. Therefore, $S_2/N$ shown in Figs. \ref{fig-4}(a) -- \ref{fig-4}(d) are slightly larger (smaller) than the curvature energy per vertex on the boundary (internal) vertices. However, this deviation is negligible in the thermodynamic limit, because the ratio $L_2/N^\prime$, which is the total number of vertices on the boundaries over the total number of internal vertices, becomes smaller and smaller as the surface size $N$ increases.

The errors $\delta_{S_2}$ of $S_2$ shown as the error bars in the figures are the so-called the standard errors and are defined as follows \cite{Janke-SASDCEE-2002}:  Let $\{S_{2i}\}(i=1,...,N_{\rm tot})$ be a sequence of MC data of $S_2$ obtained at every $1000$ MCS, and $N_{\rm tot}$ be the total number of data in the sequence. The total number of MCS performed is thus given by $1000\!\times\! N_{\rm tot}$. The series $\{S_{2i}\}$ is split into $n_b$ sub-series, and $S_{2I}$ denotes the mean value of the $I$-th sub series. Thus, we have a series of mean values $\{S_{2I}\}(I=1,...,n_b)$. Then, $\delta_{S_2}$ is defined by the standard deviation of $\{S_{2I}\}$ such that $\delta_{S_2}\!=\!\sqrt{\sum_{I=1}^{n_b}(S_{2I}\!-\!S_2)^2/n_b}$, where $S_2$ denotes the mean value. In contrast to the standard deviation of the series $\{S_{2i}\}$, $\delta_{S_2}$ decreases if $n_b$ is sufficiently large such that $\{S_{2I}\}$ are statistically independent. In the analysis of data, $n_b$ is fixed to $n_b\!=\!1\times 10^4$ in this paper, then the total number of MCS in a sub-series is $1\times 10^7$. The error bars on $C_{S_2}$ are also the standard errors.

\begin{figure}[hbt]
\centering
\includegraphics[width=14.5cm]{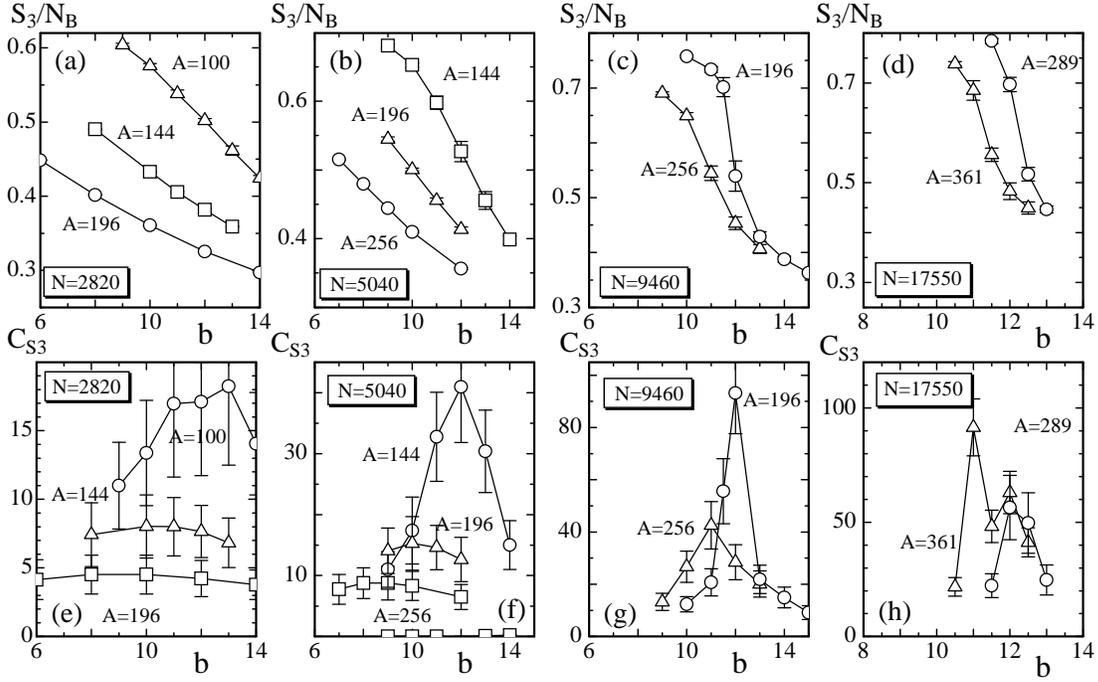}
\caption{The two-dimensional bending energy $S_3/N_B$ vs. $b$ obtained on the surfaces of size (a) $N\!=\!2820$, (b) $N\!=\!5040$, (c) $N\!=\!9460$, and (d) $N\!=\!17550$, and (e)--(h) the variance $C_{S_3}$ vs. $b$ on the same sized surfaces. }
\label{fig-5}
\end{figure}
The two-dimensional bending energy $S_3$ is defined by 
\begin{equation}
\label{two-dim-bending}
S_3=\sum_{ij} (1-{\bf n}_i\cdot {\bf n}_j),
\end{equation}
where ${\bf n}_i$ is a unit normal vector of the triangle $i$, and $\sum_{ij}$ denotes the summation over the nearest neighbor triangles $i$ and $j$. The surface fluctuations can be reflected in $S_3$. Figures \ref{fig-5}(a) -- \ref{fig-5}(d) show the two-dimensional bending energy $S_3/N_B$ vs. $b$ obtained on the surfaces of size $N\!=\!2820$, $N\!=\!5040$, $N\!=\!9460$, and $N\!=\!17550$, respectively. $N_B$ denotes the total number of internal bonds, where $S_3$ is obtained. The variance $C_{S_3}\!=\! \langle \; \left( S_3 \!-\! \langle S_3 \rangle\right)^2\rangle/N^\prime$ is shown in Figs. \ref{fig-5}(e) -- \ref{fig-5}(h). We see that the discontinuity in $S_3/N_B$ in Figs. \ref{fig-5}(c) and \ref{fig-5}(d), and the discontinuity is more accurate than that in the curvature energy $S_2/N$ in Figs. \ref{fig-4}(c) and \ref{fig-4}(d). The anomalous peaks seen in $C_{S_3}$ are also more accurate than those in $C_{S_2}$. The results shown in Figs. \ref{fig-5}(a) -- \ref{fig-5}(h) strongly indicate that the model undergoes a first-order transition of surface fluctuations.

\begin{figure}[hbt]
\centering
\includegraphics[width=14.5cm]{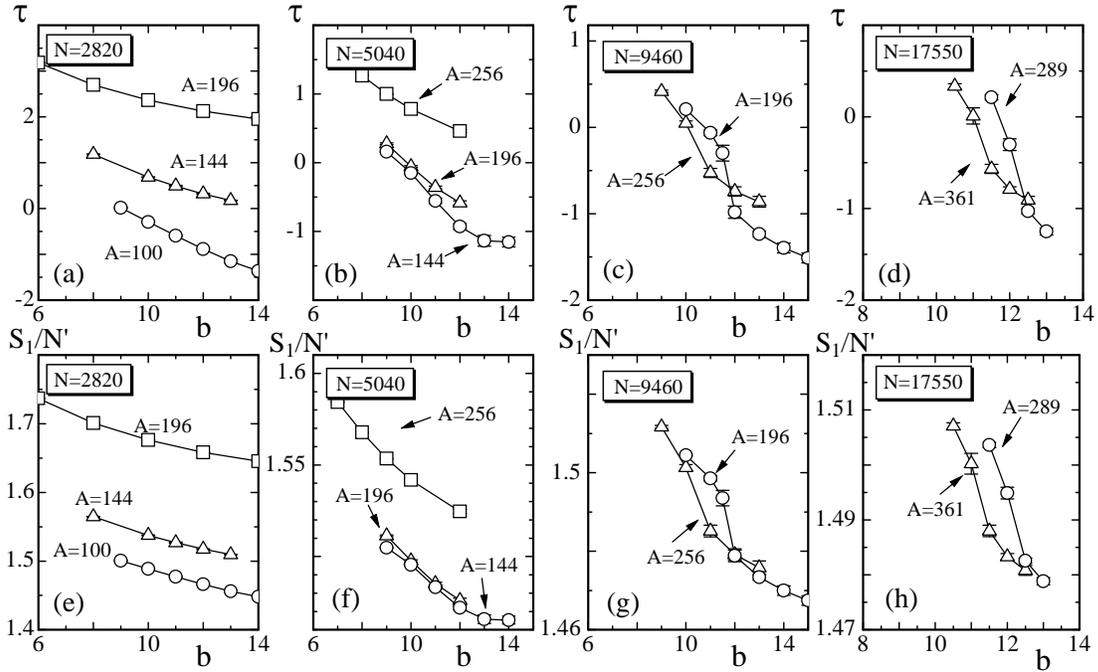}
\caption{The surface tension $\tau$ versus $b$ obtained on the surfaces of size (a) $N\!=\!2820$, (b) $N\!=\!5040$, (c) $N\!=\!9460$, and (d) $N\!=\!17550$, and (e)--(h) the Gaussian bond potential $S_1/N$ vs. $b$ on the same sized surfaces.  }
\label{fig-6}
\end{figure}
The phase transition indicated by a discontinuity of $S_2/N$ and that of $S_3/N_B$ is expected to be reflected in the surface tension $\tau$ defined in Eq. (\ref{surfacetension}). Figures \ref{fig-6}(a) -- \ref{fig-6}(d) show the surface tension $\tau$ vs. $b$, which is obtained on the surfaces of size $N\!=\!2820$, $N\!=\!5040$, $N\!=\!9460$, and $N\!=\!17550$, respectively. The surface tension discontinuously changes at the transition point $b_c$, where $S_2/N$ and $S_3/N_B$ discontinuously change. The gap seen in $\tau$ indicates that the transition is characterized by a difference in the macroscopic surface tension.

The surface tension is calculated by using the value of the Gaussian bond potential $S_1$, therefore it is expected that $S_1$ also discontinuously changes. Figures \ref{fig-6}(e) -- \ref{fig-6}(h) show the Gaussian bond potential $S_1/N^\prime$ vs. $b$ obtained on the surfaces of size $N\!=\!2820$, $N\!=\!5040$, $N\!=\!9460$, and $N\!=\!17550$, respectively. We see the expected gap of $S_1/N^\prime$ in Figs. \ref{fig-6}(g) and \ref{fig-6}(h). We should note that $S_1/N^\prime\!=\!1.5$ is satisfied in the case of a model without boundaries such as the spherical surface model because of the scale invariant property of the partition function. The reason why $S_1/N^\prime$ deviates from $S_1/N^\prime\!=\!1.5$ shown in Figs. \ref{fig-6}(e) -- \ref{fig-6}(h) is because of the constraint on $X$ imposed by the boundaries.

\begin{figure}[hbt]
\centering
\includegraphics[width=10cm]{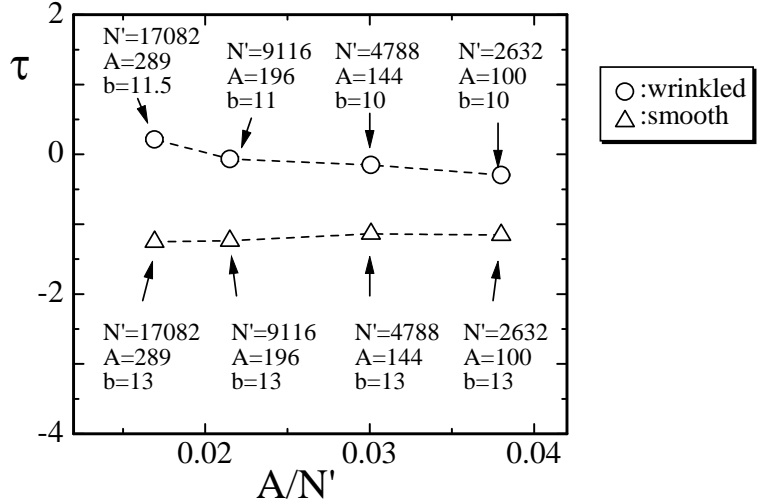}
\caption{Linear plots of $\tau$ vs. $A/N^\prime$ obtained in the smooth phase ($\bigcirc$) and in the wrinkled phase ($\triangle$).  }
\label{fig-7}
\end{figure}
The surface tension $\tau$ is expected to scale according to
\begin{equation}
\label{scaling}
\tau +\tau_0= (A/N^\prime)^\nu,
\end{equation}
where $\tau_0$ is a constant that depends on $b$ \cite{AMBJORN-NPB1993,WHEATER-JP1994}. The area $A$ was defined in the previous section by using the parameter $r$ such that $A\!=\!(rL_1)^2$, where $L_1$ is the distance between the boundaries shown in Fig. \ref{fig-1}. Since $\pi L_1$ can be identified with $L_2$, then $N^\prime$ is given by $N^\prime\!=\!L_1L_2\!-\!2L_2\!=\!\pi L_1(L_1\!-\!2)\!\simeq\!\pi L_1^2$. Thus we have $A/N^\prime\!=\!r^2/\pi$, and then we should vary $r$ in order to plot $\tau\!+\!\tau_0$ against $A/N^\prime$. It is possible to vary $r$ with fixing $N^\prime$, however, we vary $r$ by changing $N^\prime$ because we are interested in the size dependence of $\tau$. 

Since $\tau_0$ is unknown, we show in Fig. \ref{fig-7} the surface tension $\tau$ vs. $A/N^\prime$ in the wrinkled phase ($\bigcirc$) and in the smooth phase ($\triangle$). The data are obtained on four types of surfaces  $(N^\prime,A)\!=\!(17082,289)$, $(N^\prime,A)\!=\!(9116,196)$, $(N^\prime,A)\!=\!(4788,144)$ and  $(N^\prime,A)\!=\!(2632,100)$, which correspond to the ratios $r\!=\!\sqrt{289}/75\!\simeq\!0.227$, $r\!=\!\sqrt{196}/55\!\simeq\!0.255$,  $r\!=\!\sqrt{144}/40\!=\!0.3$,  and $r\!=\!\sqrt{100}/30\!\simeq\!0.33$, where $r$ is given by $r\!=\!\sqrt{A}/L_1$. The tension $\tau$ in the smooth phase is obtained at $b\!=\!13$, while $\tau$ in the smooth phase is at the range $10\leq b\leq 11.5$, where the surface is expected to be sufficiently wrinkled in each $N^\prime$.
 
From the results shown in the figure, we find that $\tau$ remains almost constant both in the smooth and in the wrinkled phases in the range $0.016\leq A/N^\prime \leq 0.038$. Moreover, $\tau$ remains almost zero in the wrinkled phase while it remains negative finite in the smooth phase.  Although the phase transition is of first-order, it is not so strong, and then the gap of the surface tension is not so clear. For this reason, the results shown in Fig. \ref{fig-7} seem not so accurate. Nevertheless, we can conclude from the results that the surface tension $\tau$  discontinuously changes at the transition point, and the value of $\tau$ in each phase  remains constant, which is independent of $A/N^\prime$ in the range $0.016\leq A/N^\prime \leq 0.038$ at sufficiently large surfaces.

\begin{figure}[hbt]
\centering
\includegraphics[width=10cm]{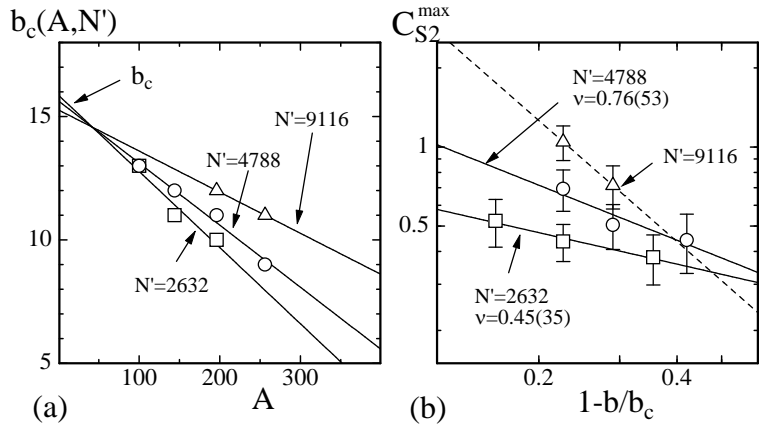}
\caption{(a) Linear plots of $b_c(A,N^\prime)$ vs. $A$ obtained on the surface of size $N^\prime \!=\!2632$, $N^\prime \!=\!4788$ and $N^\prime \!=\!9116$, and (b) log-log plots of $C_{S_2}^{\rm max}$ against $1\!-\!b/b_c$.  }
\label{fig-8}
\end{figure}
Finally, we comment on a scaling property of the specific heat $C_{S_2}$ and the strength of the phase transition. As we see in Figs. \ref{fig-4}(e)--\ref{fig-4}(h), the specific heat has the peak  $C_{S_2}^{\rm max}$ at $b_c(A,N^\prime)$, which is the critical bending rigidity and depends on both of the area $A$ and the surface size $N^\prime$. We show in Fig. \ref{fig-8}(a) linear plots of $b_c(A,N^\prime)$ against $A$, which correspond to the data in Figs. \ref{fig-4}(e), \ref{fig-4}(f) and \ref{fig-4}(g). The straight lines in Fig. \ref{fig-8}(a) are drawn by the linear fit of the data, then we have $b_c\!\simeq\!15.5$, which is seen to be independent of $N^\prime$. Therefore,  $b_c\!\simeq\!15.5$ is considered to be the critical bending rigidity of the model in the limit of $N^\prime\!\to\!\infty$. We should note that the same value of $b_c$ can also be obtained from $C_{S_3}^{\rm max}$ in Figs. \ref{fig-5}(e)--\ref{fig-5}(g). By using this value of $b_c$, we plot $C_{S_2}^{\rm max}$ against $1\!-\!b/b_c$ in a log-log scale in Fig. \ref{fig-8}(b). The straight lines in  Fig. \ref{fig-8}(b) were drawn by fitting the data to $C_{S_2}^{\rm max} \sim (1\!-\!b/b_c)^{-\nu}$; the exponent $\nu$ should be $\nu\!=\!1$ if the transition is of first-order.  We have $\nu\!=\!0.45\pm0.35$ on the $N^\prime \!=\!2632$ surface and $\nu\!=\!0.76\pm0.53$ on the $N^\prime \!=\!4788$ surface. We find that $\sigma$ increases as the surface size $N^\prime$ increases. $\nu$ seems less acurate on the surface of $N^\prime \!=\!9116$ and is not obtained because the total number of data ($=2$) is too small for the fitting. Nevertheless, the results shown in Fig. \ref{fig-8}(b) are considered to be consistent with the first-order transition.

\section{Summary and conclusions}
In this paper we study a triangulated tubular surface model with a pair of fixed one-dimensional boundaries by using the canonical Monte Carlo simulation technique. The model is defined by the Hamiltonian that is a linear combination of the Gaussian bond potential $S_1$ and the intrinsic curvature energy $S_2$. The model is known to have first-order transitions on the surfaces without boundaries and on those with point boundaries \cite{KOIB-EPJB2004,KOIB-JSAT2006-1}. However, it is non-trivial whether the model undergoes a phase transition with the one-dimensional boundaries. The surface shape of the model in this paper is almost fixed by the one-dimensional boundaries, and hence, the transition of surface fluctuations is expected to be influenced by such boundary conditions because the transition is always accompanied by the collapsing transition, which is a transition of shape transformations. Therefore, special attentions are paid on whether the model undergoes a phase transition and how the phase structure is reflected in the macroscopic surface tension $\tau$. In order to compute $\tau$, we assume several values of the projected area $A$ of the tubular surface of a given size $N^\prime$. The surface tension is expected to scale according to $\tau+\tau_0=(A/N^\prime)^\nu$ with a constant $\tau_0$ in the limit of $N^\prime\to\infty$ and $A\to\infty$ while $A/N^\prime$ is fixed \cite{AMBJORN-NPB1993,WHEATER-JP1994}. However, we are interested in the dependence of $\tau$ on $A/N^\prime$ rather than the scaling behavior because no information of $\tau_0$ is obtained. The assumed values of $A/N^\prime$ are in the range $0.016\leq A/N^\prime \leq 0.038$. We use the surfaces of size up to $N^\prime\!=\!17082$, which excludes the total number vertices on the fixed boundaries.

Firstly, we find that the model undergoes a first-order transition of surface fluctuations, which separates the smooth phase from the wrinkled phase. Both of the potential $S_1$ and the intrinsic curvature energy $S_2$ discontinuously change at the transition point $b_c$ on relatively large sized surfaces in the range $0.016\leq A/N^\prime \leq 0.038$. The two-dimensional bending energy $S_3$, which is not included in the Hamiltonian, has also a gap at the transition point $b_c$, and this indicates that the smoothness of the surface discontinuously changes at $b_c$. Secondly, the surface tension $\tau$ is also found to change discontinuously at the transition point $b_c$. The value of $\tau$ is constant in the smooth phase and it is almost zero in the wrinkled phase at $b(\not=\!b_c)$ close to the transition point in the range $0.016\leq A/N^\prime \leq 0.038$ at sufficiently large $N^\prime$, and hence the gap of $\tau$ remains finite at the transition point. Finally, we find by   finite-size scaling anaysis that the specific heat $C_{S_2}^{\rm max}$ scales according to an expected relation. The results are consistent with the first-order transition.

It is interesting to study whether the transition is seen and reflected in the surface tension in the same model on fluid surfaces. It is also interesting to study connections between the surface tension and the phase transitions in the other type of surface models.  

\section*{Acknowledgment}
This work is supported in part by a Grant-in-Aid for Scientific Research from Japan Society for the Promotion of Science.




\section*{References}
\begin{thebibliography}{00}





\bibitem{SEIFERT-LECTURE2004}
U. Seifert, Fluid Vesicles, 2004 in {Lecture Notes: Physics Meets Biology. From Soft Matter to Cell Biology.}, 35th Spring Scool, Institute of Solid State Research, Forschungszentrum J${\ddot {\rm u}}$lich.

\bibitem{Yoshikawa-FEBS2003}
K. Akiyoshi, A. Itaya, S. M. Nomura, N. Ono and K. Yoshikawa, 2003 FEBS Lett. {\bf 534} 33.

\bibitem{Hotani-JMB1984}
H. Hotani, 1984 J. Mol. Biol., {\bf 178} 113.

\bibitem{NELSON-SMMS2004-1}
D. Nelson, 2004 in {Statistical Mechanics of Membranes and Surfaces, Second Edition}, edited by  D. Nelson, T.Piran, and S.Weinberg, (World Scientific) p.1. 

\bibitem{GK-SMMS2004}
G. Gompper, and D.M. Kroll, 2004 in {Statistical Mechanics of Membranes and Surfaces, Second Edition}, edited by  D. Nelson, T.Piran, and S.Weinberg, (World Scientific), p.359. 

\bibitem{Bowick-PREP2001}
M. Bowick and A. Travesset,  2001 Phys. Rep. {\bf 344} 255.

\bibitem{Gompper-Schick-PTC-1994}
G. Gompper and M. Schick, 1994 \textit{Self-assembling amphiphilic systems}, In
\textit{Phase Transitions and Critical Phenomena 16}, C. Domb and J.L. Lebowitz, Eds. (Academic Press) p.1.

\bibitem{HELFRICH-1973}
 W. Helfrich, 1973 Z. Naturforsch, {\bf 28c} 693.

\bibitem{POLYAKOV-NPB1986}
 A.M. Polyakov, 1986 Nucl. Phys. B {\bf 268} 406.

\bibitem{KLEINERT-PLB1986}
 H. Kleinert, 1986 Phys. Lett. {\bf 174B} 335.

\bibitem{Peliti-Leibler-PRL1985}
 L. Peliti and S. Leibler, 1985 Phys. Rev. Lett. {\bf 54} 1690.

\bibitem{DavidGuitter-EPL1988}
 F. David and E. Guitter, 1988 Europhys. Lett,  {\bf 5} 709.

\bibitem{PKN-PRL1988}
M. Paczuski, M. Kardar, and D. R. Nelson, 1988 Phys. Rev. Lett. {\bf 60} 2638.

\bibitem{KANTOR-NELSON-PRA1987}
 Y. Kantor and  D.R. Nelson, 1987 Phys. Rev. A {\bf 36} 4020.

\bibitem{WHEATER-NPB1996}
J.F. Wheater, 1996 Nucl. Phys. B {\bf 458} 671.

\bibitem{KD-PRE2002}
J-P. Kownacki and H. T. Diep, 2002 Phys. Rev. E {\bf 66} 066105.

\bibitem{KOIB-PRE-20045-NPB-2006}
H. Koibuchi, N. Kusano, A. Nidaira, K. Suzuki, and M. Yamada, 2004 Phys. Rev. E {\bf 69} 066139; \\
 H. Koibuchi and T. Kuwahata, 2005 Phys. Rev. E {\bf 72} 026124; \\ 
 I. Endo and H. Koibuchi, 2006  Nucl. Phys. B {\bf 732 [FS]} 426.

\bibitem{KOIB-EPJB2004}
H. Koibuchi, N. Kusano, A. Nidaira, Z. Sasaki, and K. Suzuki, 2004 Euro. Phys. J. B {\bf 42} 561.

\bibitem{KOIB-PLA2005}
M. Igawa, H. Koibuchi, and M. Yamada, 2005 Phys. Lett. A {\bf 338} 433.

\bibitem{KOIB-PLA2006}
I. Endo and H. Koibuchi, 2006 Phys. Lett. A {\bf 350} 11.

\bibitem{KOIB-PRE2004}
H. Koibuchi, Z. Sasaki, and K. Shinohara, 2004 Phys. Rev. E {\bf 70} 066144.

\bibitem{KOIB-EPJB2007-2}
H. Koibuchi, 2007 Euro. Phys. J. B {\bf 59} 55.

\bibitem{KOIB-PRE2007}
H. Koibuchi, 2007 Phys. Rev. E {\bf 75} 051115; 2007 Phys. Rev. E {\bf 76} 061105.

\bibitem{KOIB-EPJB2007-3}
H. Koibuchi, 2007 Euro. Phys. J. B {\bf 59} 405.

\bibitem{KOIB-PLA2007}
H. Koibuchi, 2007 Phys. Lett. A {\bf 371} 278.

\bibitem{NG-PRL2004}
H. Noguchi and G. Gompper, 2004 Phys. Rev. Lett. {\bf 93} 258102.
\bibitem{AMBJORN-NPB1993}
 J. Ambjorn, A. Irback, J. Jurkiewicz, and B. Petersson, 1993 Nucl. Phys. B {\bf 393} 571.

\bibitem{WHEATER-JP1994}
 J.F. Wheater, 1994 J. Phys. A Math. Gen. {\bf 27} 3323.

\bibitem{KOIB-PLA2004-2}
H. Koibuchi, 2004 Phys. Lett. A {\bf 332} 141.

\bibitem{KOIB-EPJB2005}
H. Koibuchi, 2005 Euro. Phys. J. B {\bf 45} 377.

\bibitem{KOIB-JSAT2006-1}
H. Koibuchi, 2006 J. Stat. Mech., P05008.

\bibitem{KOIB-JSAT2006-2}
S. Obata, M. Egashira, T. Endo, and H. Koibuchi, 2006 J. Stat. Mech.,  P11016.
\bibitem{Janke-SASDCEE-2002}
Wolfhard Janke, 2002 {\it Statistical Analysis of Simulations: Data Correlations and Error Estimation} in {Quantum Simulations of Complex Many-Body Systems: From Theory to Algorithms, Lecture Notes}, J. Grotendorst, D. Marx, and A. Muramatsu (Eds.), (John von Neumann Institute for Computing, J${\ddot {\rm u}}$lich), NIC Series, Vol. {\bf 10}, pp.423--445.



\end{thebibliography}
\end{document}